\documentclass[aps,prd,showpacs,nofootinbib,amssymb,preprint]{revtex4}
\pdfoutput=1
\usepackage{amssymb,graphicx,color,ulem,amsmath}

\def\roughly#1{\mathrel{\raise.3ex\hbox
{$#1$\kern-.75em\lower1ex\hbox{$\sim$}}}}

\begin{document}

\title{The 17 MeV Anomaly in Beryllium Decays and $U(1)$ Portal to Dark Matter}

\author{Chian-Shu~Chen$^{a}$\footnote{chianshu@gmail.com}, Guey-Lin Lin$^{b}$\footnote{glin@cc.nctu.edu.tw}, Yen-Hsun Lin$^{b}$\footnote{chris.py99g@g2.nctu.edu.tw},  Fanrong Xu$^{c}$\footnote{fanrongxu@jnu.edu.cn}}
  \affiliation{$^{a}$Department of Physics, Tamkang University, New Taipei City 25137, Taiwan.\\
$^{b}$Institute of Physics, National Chiao Tung University, Hsinchu 30010, Taiwan\\
$^{c}$Department of Physics, Jinan University, Guangzhou 510632, P.R. China}

\date{Draft \today}

\begin{abstract}
The experiment of Krasznahorkay \textit{et al} observed the transition of a $\rm{^{8}Be}$ excited state to its ground state and accompanied by an emission of $e^{+}e^{-}$ pair with 17 MeV invariant mass. This 6.8$\sigma$ anomaly can be fitted by a new light gauge boson. We consider the new particle as a $U(1)$ gauge boson, $Z'$, which plays as a portal linking dark sector and visible sector. 
In particular, we study the new $U(1)$ gauge symmetry as a hidden or non-hidden group separately. The generic hidden $U(1)$ model, referred to as dark $Z$ model,  is excluded by imposing various experimental constraints. On the other hand, a non-hidden $Z'$ is allowed due to additional interactions between $Z'$ and Standard Model fermions. We also study the implication of the dark matter direct search on such a scenario. We found the search for the DM-nucleon scattering cannot probe the parameter space that is allowed by $\rm{^{8}Be}$-anomaly for the range of DM mass above 500 MeV. However, the DM-electron scattering for DM between 20 and 50 MeV can test the underlying $U(1)$ portal model using the future Si and Ge detectors with $5e^{-}$ threshold charges.  
\end{abstract}

\maketitle

\section{Introduction}\label{sec:introduction}
Recently, the experiment of Krasznahorkay \textit{et al}. studied the decays of a $\rm{^{8}Be}$ excited state to its ground state and found a bump in both the opening angle and invariant mass distributions of $e^{+}e^{-}$ pairs produced in the transitions~\cite{Krasznahorkay:2015iga}. This 6.8$\sigma$ deviation from the expectation can be fitted by the production of a new particle $\rm{X}$ of mass around 17 MeV in the transition $\rm{^{8}Be^*} \rightarrow \rm{^{8}Be}~X$ and the subsequent decay of $\rm{X}$ into electron-positron pair. Although the excess can be due to unknown nuclear effects, the existence of a new particle was investigated in ~\cite{Feng:2016jff,Gu:2016ege,Chen:2016dhm,Liang:2016ffe,Chen:2016kxw,Feng:2016ysn,Kitahara:2016zyb,Ellwanger:2016wfe,Chu:2016pew,Jia:2016uxs} from different aspects. It is shown that the proton coupling to $\rm{X}$ is suppressed, or the $\rm{X}$ boson is $protophobic$, in order to alleviate various experimental constraints for a new light gauge boson~\cite{Feng:2016jff,Feng:2016ysn}. Some anomaly free phenomenological models are also proposed in Refs.~\cite{Gu:2016ege,Feng:2016ysn}. Furthermore, simple models involving new scalar or pseudo-scalar particles are excluded as shown in Ref.~\cite{Feng:2016ysn}.

In this paper, we take the new particle as a $Z'$ gauge boson corresponding to a new $U(1)_d$ gauge symmetry. The charge $d$ can be a completely hidden quantum number governing the dark sector or a certain combination of Standard Model (SM) quantum numbers. Hence the $Z'$ boson plays as a portal between the dark sector and visible sector.  A simple dark photon model containing two parameters $(\varepsilon, m)$ has been ruled out by confronting
various experiments both in quark sector and lepton sector \cite{Feng:2016jff}. 
The dark $Z$ model, however,  has one more parameter $\varepsilon_Z$ which is induced by
 the mass mixing between $Z$ and $Z'$. This mixing allows interactions between $Z'$ and fermions including neutrinos.
Then it is a natural and interesting topic to discuss the possibility of this more generic model.  
In the following sections, we will show that the hidden gauge boson couples to proton (neutron) and electron (neutrino) in equal strengths. 
Due to the above coupling relations, we will present explicitly the incompatiblity between the measurement of Krasznahorkay \textit{et al.}~\cite{Krasznahorkay:2015iga} and TEXONO $\nu-$electron scattering experiments~\cite{Deniz:2009mu}. 
Since the generic hidden $U(1)$ model is disfavored, we also consider non-hidden $U(1)$ portal models for further phenomenological studies. We shall demonstrate that such models are also disfavored by the existing DM direct search data for DM mass above 500 MeV. To probe DM lighter than 500 MeV in non-hidden models, we propose direct searches based upon DM-electron scatterings.\footnote{A similar idea for scalar and vector DM was investigated in Ref.~\cite{Jia:2016uxs}. Here we focus on fermionic DM with a p-wave contribution to DM annihilation cross section during thermal freeze-out epoch.}. 


We organized the paper as follows : in section II we describe the generic $Z'$ Lagrangian and take the $Z'$ boson as the DM force mediator;  in section III we derive the constraints for the model parameters from the $\rm{^{8}Be}$ transition and other experiments, in particular TEXONO experiment excludes the simple dark $Z$ model; in section IV we investigate the implications of DM direct search with DM-nucleus scatterings and discuss the sensitivities of future Si and Ge detectors to DM-electron scattering in the framework of non-hidden $U(1)$ portal model with the WIMP mass between 20 MeV and 500 MeV; we then conclude in Section \ref{sec:summary}.   

\section{The framework}

\subsection{The generic hidden $U(1)$ model}
We assume that the dark sector interacts among themselves and links the SM particles via a hidden abelian $U(1)_{d}$ gauge symmetry. The connection between visible and dark sector is established by 
the  $U(1)_{d}$ gauge boson $Z'$ through its kinetic mixing with the SM 
$U(1)$ gauge boson $B$, 
\begin{equation}
\mathcal{L}_{{\rm gauge}} =-\frac14 B_{\mu\nu}B^{\mu\nu}+ \frac12\frac{\varepsilon}{\cos\theta_W}Z'_{\mu\nu}B^{\mu\nu}
-\frac14 Z'_{\mu\nu}Z'^{\mu\nu}, 
\end{equation}
where  $\varepsilon$ characterizes the kinetic mixing. 
After spontaneous symmetry breaking 
(SSB) $W$ and $Z$ bosons get their masses while photon stays massless in SM. 
 The mass of $Z'$, which is quite model dependent,  can be generated either 
 from St\"uckelberg mechanism (one example can be found in \cite{Feng:2014cla}) 
or by applying Higgs mechanism (for example see \cite{Davoudiasl:2012ag}). 
Generically in addition to kinetic mixing interaction as shown eq.(1),
there is a mass mixing term between $Z'$ and $Z$.
For simplicity, here we do not distinguish the $Z'$ gauge eigenstate and its mass state (both states 
are related by
 several rotations) and 
keep in mind hereafter $Z'$ will be in mass eigenstate. 
Without involving detailed information of UV complete models, 
one can always obtain effective interactions among $Z'$ and SM fermions 
by diagonalizing the kinetic mixing and mass mixing terms,
%

\begin{equation}
\mathcal{L}_{\textrm{visible}} = - \left(\varepsilon_{\gamma}eJ^{\mu}_{{\rm em}} + \varepsilon_{Z}\frac{g}{2c_{W}}J^{\mu}_{\rm NC}\right)Z'_{\mu} 
\end{equation}
where $\varepsilon_\gamma=\varepsilon$, and $\varepsilon_Z$
is  certain combination of $\varepsilon$ and 
rotation angles, which depends on UV complete theory and mass generation mechanism. The electromagnetic current and weak neutral current are 
\begin{equation}
J^{\mu}_{{\rm em},f} = Q_{f}\bar{f}\gamma^{\mu}f \quad {\rm and} \quad J^{\mu}_{{\rm NC},f} = (T_{3f} - 2Q_{f}s^2_{W})\bar{f}\gamma^{\mu}f - T_{3f}\bar{f}\gamma^{\mu}\gamma_5f
\end{equation}
respectively, where $f$ stands for the fermions with corresponding electric charge $Q_{f}$, isospin $T_{3f} = \pm\frac{1}{2}$. More details about this model can be found in~\cite{Davoudiasl:2012ag, Xu:2015wja}. Furthermore, we assume 
the interaction in dark sector is
\begin{equation}
\mathcal{L}_{\textrm{dark}} = e_{d}\bar{\chi}\gamma^{\mu}\chi Z'_{\mu} ,
\end{equation}
where 
the DM $\chi$ is assumed to be Dirac fermion carrying a charge $e_{d}$ under the hidden gauge symmetry. 

\subsection{$Z' \rightarrow e^{+}e^{-}$ and $Z' \rightarrow \nu\bar{\nu}$}
In fitting to the measurement by Krasznahorkay \textit{et al}., one obtains the mass of the light boson to be~\cite{Krasznahorkay:2015iga} 
\begin{equation}
M_{Z'} = 16.7\pm0.35(\rm stat.)\pm0.5(\rm sys.)~{\rm MeV}.
\end{equation}
Therefore, the $Z'$ considered in our framework can only decay into $e^{+}e^{-}$ and $\nu\bar{\nu}$. The corresponding decay widths are given by  
\begin{equation}\label{GammaE}
\Gamma(Z' \rightarrow e^{+}e^{-}) = \alpha_{\rm em}(a^2_{e} + b^2_e)\frac{M^2_{Z'} + 2m^2_{e}}{3M_{Z'}}\sqrt{1 - \frac{4m^2_{e}}{M^2_{Z'}}} 
\end{equation}
and
\begin{equation}\label{Gammanu}
\Gamma(Z' \rightarrow \nu_{i}\bar{\nu}_{i}) = \alpha_{\rm em}(a^2_{\nu} + b^2_{\nu})M_{Z'}
\end{equation}
respectively\footnote{One should bear in mind that a dark sector decay channel may also open for the corresponding final state particle mass is kinematically allowed. Hence the branching ratio of $Z' \rightarrow e^+e^-$ could be a tuning parameter.}. We have summed up three flavors of neutrino and the parameters $a_{f}$ and $b_{f}$ are defined as 
\begin{equation}
a_{f} = Q_{f}\varepsilon_{\gamma} + \frac{T_{3f} - 2Q_{f}s^2_{W}}{2c_{W}s_{W}}\varepsilon_{Z} \quad {\rm and} \quad b_{f} = -\frac{T_{3f}}{2c_{W}s_{W}}\varepsilon_{Z}.
\end{equation}
Numerically, we have $a_{e} =-\varepsilon_{\gamma}-0.05\varepsilon_{Z}$, $b_{e} = -0.6\varepsilon_{Z}$ and $a_{\nu}(b_{\nu}) = - (+)0.6\varepsilon_{Z}$ respectively. 

\section{The constraints in generic hidden $U(1)$ model}

\subsection{The explanation to ${\rm {^8Be^*}} \rightarrow {\rm {^8Be}}~Z'$  and other experimental constraints}
The dark $Z'$ interaction with nucleon can be characterized as 
\begin{equation}
- \mathcal{L}_{{\rm N}} = Z'^{\mu}(J_{\mu}^N + J_{5\mu}^{N} )
\end{equation}
with $J_{\mu}^{N} = e\varepsilon_{p}\bar{p}\gamma_{\mu}p + e\varepsilon_{n}\bar{n}\gamma_{\mu}n$ and $J_{5\mu}^{N} = e\varepsilon_{p5}\bar{p}\gamma_{\mu}\gamma_5p + e\varepsilon_{n5}\bar{n}\gamma_{\mu}\gamma_5n$ respectively. Only the vector current part contributes to the matrix element $\langle {\rm ^8Be }Z'|\mathcal{L}_{\rm N}|{\rm ^8Be^*} \rangle$ if parity is conserved. It is interesting that the couplings for proton and neutron are identical to $a_{e}$ and $a_{\nu}$ except the sign flip by opposite electric charge and weak isospin. Thus, we deduce that 
\begin{subequations}\begin{align}
\varepsilon_{p} & = 2a_u+a_d = \varepsilon_{\gamma} + 0.05\varepsilon_{Z} = -a_{e}\label{ep=ae} \\
\varepsilon_{n} & = a_u+2a_d = -0.6\varepsilon_{Z} = -a_{\nu}\label{en=anu}.
\end{align}\end{subequations}
According to Refs.~\cite{Krasznahorkay:2015iga,Feng:2016jff}, we have 
\begin{equation}
\frac{\Gamma(^8{\rm Be}^*\rightarrow ^8{\rm Be}Z')}{\Gamma(^8{\rm Be}^*\rightarrow ^8{\rm Be}\gamma)}Br(Z'\rightarrow e^+e^-) = 5.8\times10^{-6}.
\end{equation}
In the isospin symmetry limit\footnote{In general the transitions between $^8{\rm Be}^*$ and $^8{\rm Be}$ can be both isovector and isoscalar. A isospin breaking derivation for the transition is presented in Ref.~\cite{Feng:2016jff}. However, the modification is only about 20\%. Since our results are not much affected, we present the isospin symmetry limit of the transition rate for simplicity.}, we have~\cite{Feng:2016jff} 
\begin{equation}\label{constraint1}
\frac{\Gamma(^8{\rm Be}^*\rightarrow ^8{\rm Be}Z')}{\Gamma(^8{\rm Be}^*\rightarrow ^8{\rm Be}\gamma)} = (\varepsilon_{p} + \varepsilon_{n})^2\left[1 - (\frac{M_{Z'}}{18.15~\rm MeV})^2\right]^{3/2}.
\end{equation}
The branching ratio of 
$Z' \rightarrow e^+e^-$ in our scenario can be deduced from Eq.~(\ref{GammaE}) and Eq.~(\ref{Gammanu}), 
\begin{equation}
Br(Z'\rightarrow e^+e^-) \approx 
\frac{a_e^2+b_e^2}{(a_e^2+b_e^2)+3(a_\nu^2+b_\nu^2)}
=\frac{\varepsilon_p^2+\varepsilon_n^2}{\varepsilon_p^2+7\varepsilon_n^2},
\end{equation}
where we have applied the relations $|b_e|=|b_\nu|=|a_\nu|=|\varepsilon_n|$, $|a_e|=|\varepsilon_p|$,
which particularly hold in dark $Z$ model.

\begin{figure}
\begin{centering}
\includegraphics[width=0.45\textwidth]{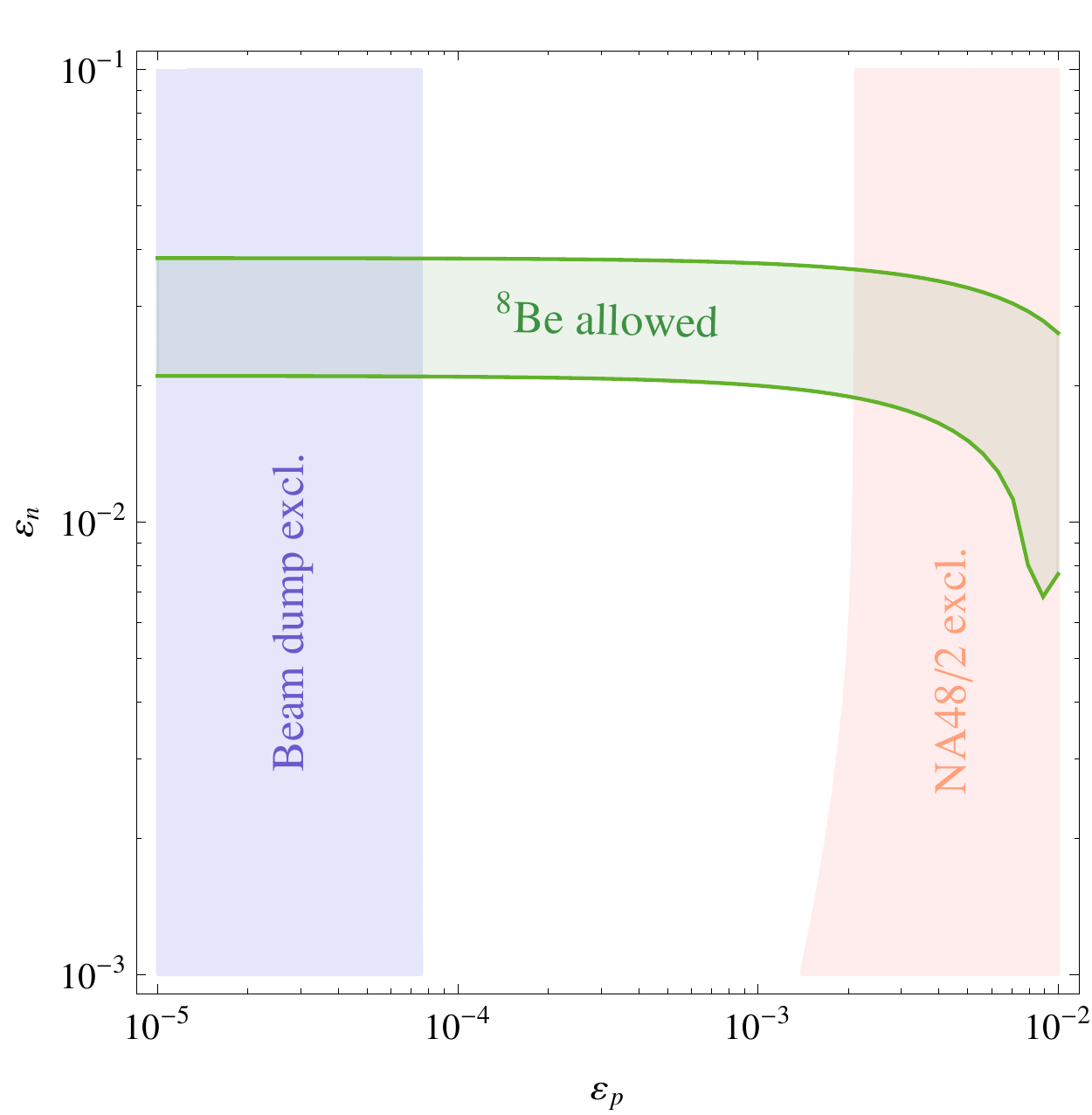}
\par\end{centering}
\caption{\label{fig:parameterspace}The allowed parameter space on $(\varepsilon_{p}, \varepsilon_{n})$ or $(a_{e}, a_{\nu})$ plane in generic hidden $U(1)$ model. E141 and NA48/2 exclusion regions are indicated. The green band is the allowed parameter space resulting from $^8{\rm Be}$ anomaly with the error of $Z'$ mass taken into account. 
The allowed narrow band is however incompatible with the TEXONO $\nu-e$ scattering experiment~\cite{Deniz:2009mu}.}
\end{figure}

In addition, as pointed in Ref.~\cite{Feng:2016jff} and constraints summarized in Ref.~\cite{Alexander:2016aln}, two most severe bounds from the search of dark photon in the relevant mass range are obtained by NA48/2~\cite{Batley:2015lha} and E141 experiments~\cite{Riordan:1987aw}. NA48/2 gives an upper bound requiring $\varepsilon_{{\rm max}} \lesssim 4.8\times10^{-4}$ at 90\% C.L.~\cite{Batley:2015lha}. 
\footnote{In \cite{Kahn:2016vjr} the NA48/2 constraint on $U(1)_d$ model is also applied similarly.}
We translate the constraint into our framework by interpretting the process as $\pi^0 \rightarrow Z'\gamma \rightarrow e^+e^-\gamma$. Again the branching ratio modification is taken into account, 
\begin{equation}\label{constraint2}
|\varepsilon_{p}| 
\lesssim \frac{(0.8 - 1.2)\times10^{-3}}{\sqrt{Br(Z'\rightarrow e^+e^-)}}.
\end{equation}
E141 is a electron beam dump experiment at SLAC which searches for a dark photon bremsstrahlung resulting from electrons incident on a nuclear target~\cite{Riordan:1987aw}. The experiment sets a lower bound for the coupling strength in our scenario
\begin{equation}\label{constraint3}
\frac{|a_{e}|}{\sqrt{Br(Z'\rightarrow e^+e^-)}} 
\gtrsim 2\times10^{-4}.
\end{equation}
Combining the condition of Eq.~(\ref{constraint1}) and two constraints, Eq.~(\ref{constraint2}) and Eq.~(\ref{constraint3}), we plot the allowed parameter region for $(\varepsilon_{p}, \varepsilon_{n})$ in Figure~\ref{fig:parameterspace}. The green band is the allowed region to fit the $^8{\rm Be}$ anomaly while the purple shaded and pink shaded areas are excluded by the beam dump and NA48/2 experiments, respectively. One observes that $\varepsilon_{n}$ lies within a narrow region between $10^{-1}$ and $10^{-2}$ and $\varepsilon_{p}$ is constrained in the range of $10^{-4} \sim 10^{-3}$. The allowed coupling strength of $Z'$ to proton is relatively smaller than the coupling strength of $Z'$ to neutron. Hence a $protophobic$ feature is suggested by the measurement of Krasznahorkay \textit{et al}..  

\subsection{$\nu-e$ scattering experimental constraint}
For the hidden $U(1)$ model, the same constraints also apply to $(a_{e}, a_{\nu})$ as shown in Eq.~(\ref{ep=ae}) and Eq.~(\ref{en=anu}). Therefore, the constraints from short baseline accelerator and/or reactor neutrino-electron scattering experiments must be taken into account~\cite{Deniz:2009mu,Auerbach:2001wg,Daraktchieva:2005kn}.\footnote{
An example of constraining $U(1)_{B-L}$ model by using $\nu-e$ scattering experiment can be found in  ~\cite{Bilmis:2015lja}.}
A global analysis on the nonstandard interactions that are deviated from the SM predictions is presented in Ref.~\cite{Khan:2016uon}. We take the effective Lagrangian approach by integrating the intermediate $Z'$ boson, which yields the following bounds
\begin{equation}
|(a_{e} - a_{\nu})a_{\nu}| \lesssim 8\times10^{-9} \quad {\rm and} \quad |(a_{e} + a_{\nu})a_{\nu}| \lesssim 5\times10^{-9}
\end{equation}
Using Eqs.~(\ref{ep=ae}) and (\ref{en=anu}), these bounds can be translated into constraints for $\varepsilon_{p}$ and $\varepsilon_{n}$. 
These constraints are so stringent that they are incompatible with the experimental constraints we just derived in the framework of generic hidden $U(1)$ model.  

\section{Non-hidden $U(1)$ portal and DM-electron scattering process}
To accommodate the new light gauge boson indicated in $^8{\rm Be}$ anomaly as well as $U(1)$ portal scenario, we are led to consider models with non-hidden $U(1)$ gauge symmetry and MeV-scale DM\footnote{An axion-like or other scenario of $m_{\chi} < 500$~MeV are viable. We concentrate our discussion on MeV-scale DM in this paper.}. Non-hidden $U(1)$ charge suggests a certain linear combination of SM quantum number and/or other hidden charge. Phenomenologically, such models will include a new set of direct gauge-fermion couplings. The interplay between these couplings with the gauge boson mixings will modify the relations among quark and lepton couplings such as Eq.~(\ref{ep=ae}) and Eq.~(\ref{en=anu}). There are various ways of model-building to impose such non-hidden $U(1)$ gauge symmetry motivated by $^8{\rm Be}$ anomaly~\cite{Gu:2016ege,Feng:2016ysn}. In this paper, we do not intend to study these models in detail but rather assume that the couplings of $Z'$ to various fermions are not correlated. In particular, we assume the severe constraint from $\nu-e$ scattering can be alleviated\footnote{One simple example is the $U(1)_{B}$ model with $B$ the baryon number. In such a model the neutrino-$Z'$ coupling vanishes, thus the TEXONO bounds can be evaded. An anomaly free $U(1)_{B}$ model is proposed in Ref.~\cite{Feng:2016ysn}}. In this section, we first discuss various constraints on those generic non-hidden $U(1)$ models, including constraints from cosmology and constraints from DM direct search with DM-nucleus scattering. We then discuss the sensitivities of future Si and Ge detectors to DM-electron scatterings for light DM in the MeV mass range.
\subsection{Thermal freeze-out and cosmological constraints}
\begin{figure}
	\begin{centering}
		\includegraphics[width=0.21\textwidth]{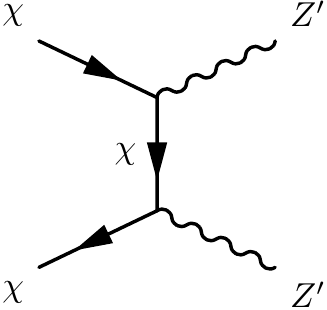} \qquad~~~
		\includegraphics[width=0.21\textwidth]{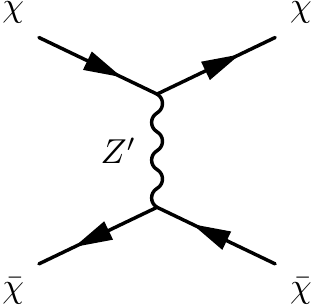} \qquad~~~
		\includegraphics[width=0.21\textwidth]{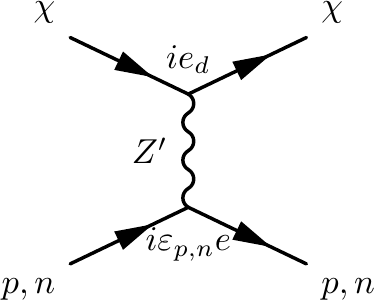}
		\par\end{centering}
	\caption{\label{fig:dm_int}\textit{left}: DM annihilation to a $Z'$ pair. This P-wave process is enhanced when the relative velocity of the annihilating DMs is small, $\langle \sigma v\rangle \propto v^{-1}$. This is due to the Sommerfeld enhancement effect. 
	\textit{middle}: DM-DM self-interaction. N-body simulation and astrophysical observations set the range of the self-interaction cross section as $0.1~ (\rm{cm^2/g}) \le \sigma/m_{\chi} \le 1 ~(\rm{cm^2/g})$. See the texts for the details. \textit{right}: DM-nucleus scattering. We calculate the DM-nucleus scattering cross section in which the vertices $e_d$ is fixed by DM relic abundance and DM self-interaction constraints while $\varepsilon_{p,n}$ are taken from the allowed region in Figure~\ref{fig:parameterspace}.}
\end{figure}
DM relic abundance requires the WIMP annihilation cross section to be around $\langle \sigma v\rangle \approx 3\times 10^{-26}~{\rm cm^3~s^{-1}}$. In our study we shall consider other cosmological constraints on $\langle \sigma v \rangle$. For cosmic microwave background (CMB) the additional injection of energy via the DM annihilations will increase the ionization fraction in the CMB anisotropy. Hence it will suppress the power spectrum at small angular scales due to the broadening of the last scattering surface, and also enhance the polarization power spectra at low multipoles due to the increasing probability of the Thomson scattering. The Planck data puts strong bounds on s-wave annihilation cross section for the DM mass ranging from sub-MeV to 100 GeV~\cite{Slatyer:2015jla}. In particular, the constraints are stringent if the fraction of electron final state is non-negligible. We note that $^8{\rm Be}$ anomaly indicates a substantial fraction of $e^+e^-$ final state in $Z'$ decays. In such a case the Planck data requires the s-wave $\langle \sigma v \rangle$ to be less than $10^{-29}\sim10^{-30}~{\rm cm^3s^{-1}}$ for MeV-scale DM~\cite{Slatyer:2015jla}. 

Here we consider the process, $\chi\chi\to Z'Z'$, as the dominant  DM annihilation channel. 
The Feynman diagram for this process is depicted on the left of Figure~\ref{fig:dm_int}. 
In the parity conservation limit, this process is mostly p-wave as pointed out in~\cite{An:2016kie}. We include the Sommerfeld enhancement factor~\cite{Iengo:2009ni,Cassel:2009wt} and find $\alpha_d \equiv  e^2_d/4\pi$, the analogous fine structure constant for $U(1)_{d}$ gauge interaction, is about $5.2\times 10^{-5}\,(m_\chi/{\rm GeV})$ to satisfy the thermal relic abundance. 
Here we already include the constraint on DM self-interaction strength, which requires $0.1~ (\rm{cm^2/g}) \le \sigma/m_{\chi} \le 1 ~(\rm{cm^2/g})$. The Feynman diagram for DM self-interaction is shown in the middle of Fig.~\ref{fig:dm_int}. The above range for $\sigma/m_{\chi}$ was proposed for resolving the discrepancies between the numerical N-body simulations using the hypothesis of collisionless cold dark matter (CCDM) and the astrophysical observations on the small structure of the universe~\cite{Spergel:1999mh}. This puzzle is  the so-called cuspy-core problem in the center regions of galaxies. The CCDM simulations~\cite{Navarro:1996gj} predicts cuspy profiles in the center regions of galaxies while  much more flatten cores are found in our Milky Way~\cite{Walker:2011zu} and other nearby dwarfts~\cite{Oh:2010ea}, and low luminous galaxies~\cite{Gentile:2004tb,KuziodeNaray:2007qi}. The other is so-called ``too-big-to-fail" problem which is referred to the sizes of subhalos. The observed Milky Way satellites are hosted by much less massive subhalos compared to sizes of those predicted by simulations~\cite{BoylanKolchin:2011de}.

In addition, the cross sections between DM and baryons  are constrained based on the linear density perturbations in cosmology~\cite{Chen:2002yh,Dvorkin:2013cea}. New interactions could transfer momentum from DM to baryon-photon fluid and modify the baryon-photon oscillations. This then affects the spectrum of CMB and the Lyman-$\alpha$ forest data from the Sloan Digital Sky Survey (SDSS). A model-independent constraint on the DM-baryon cross sections (power-law dependence on DM-baryon relative velocity) was investigated in~\cite{Dvorkin:2013cea}. Specifically, the DM-baryon cross section, which is essentially the cross section between DM and hydrogen, is parameterized as $\sigma_0v^n$ with $v$ the relative velocity between DM and baryon. The constraint on $\sigma_0$ is given for $n=0, -1, -2, {\rm and} -4$. In the next subsection, we shall compare such constraints with the DM-baryon cross section predicted by non-hidden $U(1)$ portal models. 

Finally, in order to prevent the photodissociation which can alter the light element abundances during the big bang nucleosynthesis, one requires the lifetime of $ Z'$ to be less than 1 second in the early universe~\cite{Holtmann:1998gd}. Hence the mixing parameter is constrained to be $\varepsilon_{Z} \gtrsim 6.8\times10^{-11}\times\sqrt{{\rm 17 MeV}/M_{Z'}}$~\cite{Lin:2011gj}.    
In this paper, we focus on the scenario that $m_\chi > M_{Z'}$. Hence the minimum $m_\chi$ in our study is $20\,{\rm MeV}$. 

\subsection{DM direct searches and the DM-nucleon cross section through the exchange of 17 MeV $Z'$}
Even though the non-hidden U(1) models can accommodate both Be-anomaly and $\nu-e$ scattering experiments, it is worthy of looking at the impact of DM direct searches on these models. The corresponding Feynman diagram is depicted on the right of Figure~\ref{fig:dm_int}.
In the limit of zero momentum transfer ($q^2 = 0$), the DM-nucleus scattering cross section is given by 
\begin{equation}
\sigma_{\chi \rm{A}} = \frac{16\pi\alpha_{\rm em}\alpha_{d}\mu^2_{\chi A}}{M^4_{Z'}}[\varepsilon_{p}Z + \varepsilon_{n}(A - Z)]^2
\end{equation}
where $Z$ and $A$ are proton number and mass number, respectively, 
and  $\mu_{\chi A}= m_{\chi}m_{A}/(m_{\chi} + m_{A})$ is the DM-nucleus reduced mass. In this case, DM-neutron coupling $\varepsilon_n$ gives the main contribution to $\sigma_{\chi A}$ due to the suppression of $\varepsilon_p$ compared to 
$\varepsilon_n$.
Additionally, the force mediator is light enough such that the propagator correction $M^4_{Z'}/(M^2_{Z'} + \mathbf{q}^2)^2$ due to the momentum transfer should be included in the cross section $\sigma_{\chi \rm{A}} $. We have included this correction in our calculation and 
have done the same for calculating $\sigma_{\chi n}$ (DM-neutron scattering cross section) and $\sigma_{\chi p}$ (DM-proton scattering cross section). With $v=270$ km/s the DM velocity dispersion, we have $\mathbf{q}^2\approx m_{\chi}^2v^2\approx (m_{\chi}/{\rm GeV})\cdot {\rm MeV}^2$.  
The coupling strength $\alpha_d \equiv  e^2_d/4\pi$ is fixed at $5.2\times 10^{-5}\,(m_\chi/{\rm GeV})$ due to thermal relics via p-wave contribution and the DM self-interaction constraints. The cross section $\sigma_{\chi A}$ can be evaluated by taking the parameter set  $(\epsilon_{p},\epsilon_n)$ according to the region allowed by Be-anomaly as shown in Figure~\ref{fig:parameterspace}. It is also useful to write 
$\sigma_{\chi A}=\sigma_{\chi n}(A-Z)^2(m_\chi +m_n)^2/m_n^2$, which is valid for $\varepsilon_n\gg \varepsilon_p$ and $m_A \gg m_\chi$. 

\begin{figure}
\begin{centering}
\includegraphics[width=0.5\textwidth]{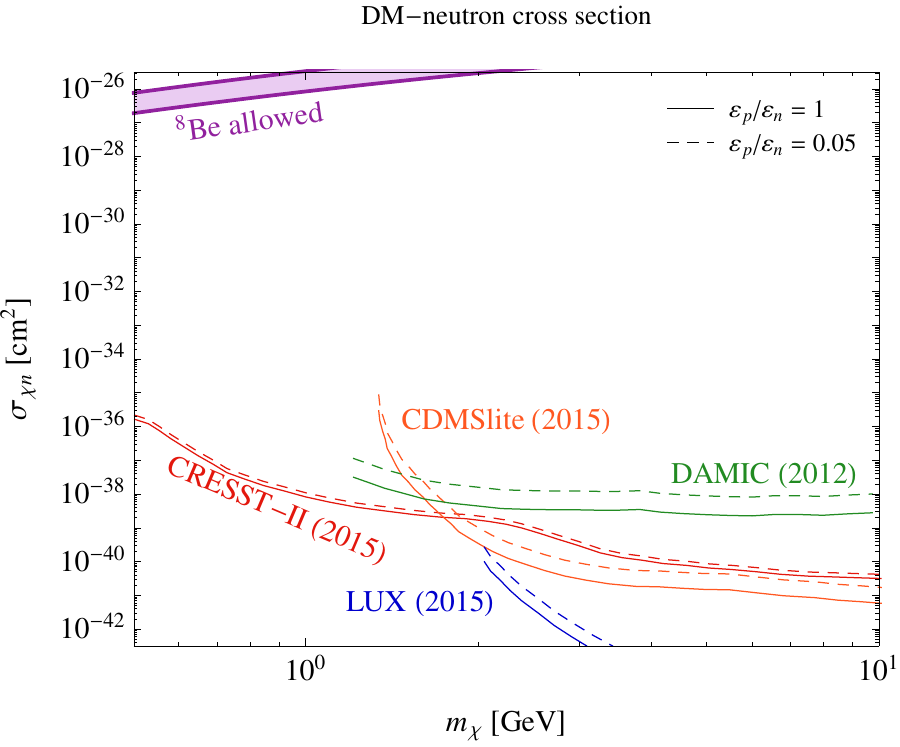}
\par\end{centering}

\caption{\label{fig:DM-nucleon_sigma}The theoretical predictions of $\sigma_{\chi n}$ from the allowed range of $(\varepsilon_{p}, \varepsilon_{n})$ parameter space is shown. The purple shaded regions are the predicted DM-neutron cross section. The exclusion lines by CRESST-II(2015)~\cite{Angloher:2015ewa}, DAMIC(2012)~\cite{Barreto:2011zu}, CDMSlite(2015)~\cite{Agnese:2015nto}, and LUX(2015)~\cite{Akerib:2015rjg} are presented for $\varepsilon_{p}/\varepsilon_{n} =1$ and $\varepsilon_{p}/\varepsilon_{n} =0.05$, respectively, where the latter ratio is the protophobic scenario favored by $^8{\rm Be}$ experiment. $\sigma_{\chi n}$ is predected to be around $10^{-26}\sim10^{-27}~{\rm cm^2}$. The attenuation length of DM with these cross section will not penetrate 1000 m rock to reach the detectors. Therefore, DM direct searches will not probe the parameter space allowed by Be-anomaly. 
}
\end{figure}

We plot in Fig.~\ref{fig:DM-nucleon_sigma} the theoretical predictions of $\sigma_{\chi n}$ for DM mass between 0.5-10 GeV. The DM-neutron cross section is around the $10^{-26}~\rm{cm}^{2}$ for $m_{\chi}=0.5$ GeV. We also show the corresponding DM direct search bounds obtained by CRESST-II(2015)~\cite{Angloher:2015ewa},  DAMIC(2012)~\cite{Barreto:2011zu}, CDMSlite(2015)~\cite{Agnese:2015nto}, and LUX~\cite{Akerib:2015rjg}\footnote{A new update result of LUX on IDM 2016 claims a constraint four
times better than the one published in 2015 during the writing of
this paper. By incorporating this new result, the exclusion region
will extend to a lower DM-nucleon cross section. Nonetheless, we still
present the LUX 2015 result as a benchmark in this work.}. Although the direct searches have set strong bounds on $\sigma_{\chi n}$, we cannot naively apply such bounds. It is because the DM cannot reach the underground detectors for 
$\sigma_{\chi n}\sim10^{-26}~\rm{cm}^2.$
This can be understood by considering a underground laboratory with 1000 m of standard rock (with $Z=11$, $A=22$ and the density $\rho=2.65~\rm{g/cm}^3$) as the overburden. For $m_{\chi}=1$ GeV, the DM with $\sigma_{\chi n}$ higher than $5.4\times10^{-30}~\rm{cm}^2$ will not be able to reach the detector since its attenuation length will be shorter than 1000~m. On the other hand, a cross section lower than $5.4\times10^{-30}~\rm{cm}^2$ will be subject to the direct search constraint. 
Here attenuation length is defined as $\Lambda=1/(n\sigma \eta)$ where $n$ is the number density of target nuclei and $\eta$ is the inelasticity of DM-rock collision, i.e., the fraction of initial DM kinetic energy transferred to the target nucleus. On average we have $\eta=m_{\chi}/m_A$. For a fixed standard rock overburden, the critical DM-neutron cross section, denoted as  $\sigma^c_{\chi n}$,  below which DM can reach 
to the detector is proportional to $m_A/m_{\chi}\cdot m_n^2/(m_{\chi}+m_n)^2$. Clearly $\sigma^c_{\chi n}$ increases as $m_{\chi}$ decreases. 

Since one cannot test the very large $\sigma_{\chi n}$ for GeV range $m_{\chi}$ by direct search, it is of interest to see whether CMB and SDSS data mentioned in the previous subsection 
can provide some constraints or not. As pointed out in Ref.~\cite{Dvorkin:2013cea}, the upper bounds on $\sigma_{\chi p}/m_{\chi}$ for a velocity-independent $\sigma_{\chi p}$ is $5.9\times10^{-27}~\rm{cm^2/GeV}$ with the combination of  CMB and Lyman-$\alpha$ data. Taking the protophobic scenario $\varepsilon_{p}/\varepsilon_{n}=0.05$ implied by Be-anomaly, such a constraint is translated into $\sigma_{\chi n}/m_{\chi} < 2.4\times10^{-24}~\rm{cm^2/GeV}$. The DM-nucleon cross section in non-hidden U(1) model discussed here is essentially velocity-independent for GeV-scale $m_{\chi}$ (the momentum transfer is less than $m_{Z'}$). Hence the Be-anomaly allowed $\sigma_{\chi n}$ is still compatible with CMB+Lyman data for GeV range $m_{\chi}$. 

We have seen that neither direct search with DM-nucleus scattering nor cosmological data can probe the predicted $\sigma_{\chi n}$ in the GeV mass range. In particular the former is due to 
the large $\sigma_{\chi n}$ that prevents DM entering the underground laboratory. We note that $\sigma^c_{\chi n}$ is proportional to $m_A/m_{\chi}\cdot m_n^2/(m_{\chi}+m_n)^2$ while the predicted 
$\sigma_{\chi n}$ is proportional to $m_{\chi}^2m_n^2/(m_{\chi}+m_n)^2$. It is clear that the chance for DM to enter the underground laboratory increases as $m_{\chi}$ decreases. This leads us to consider MeV DM which is to be detected by DM-electron scattering instead of DM-nucleus process as will be discussed later.  



\subsection{DM-electron scattering process}
\begin{figure}
	\begin{centering}
		\includegraphics[width=0.5\textwidth]{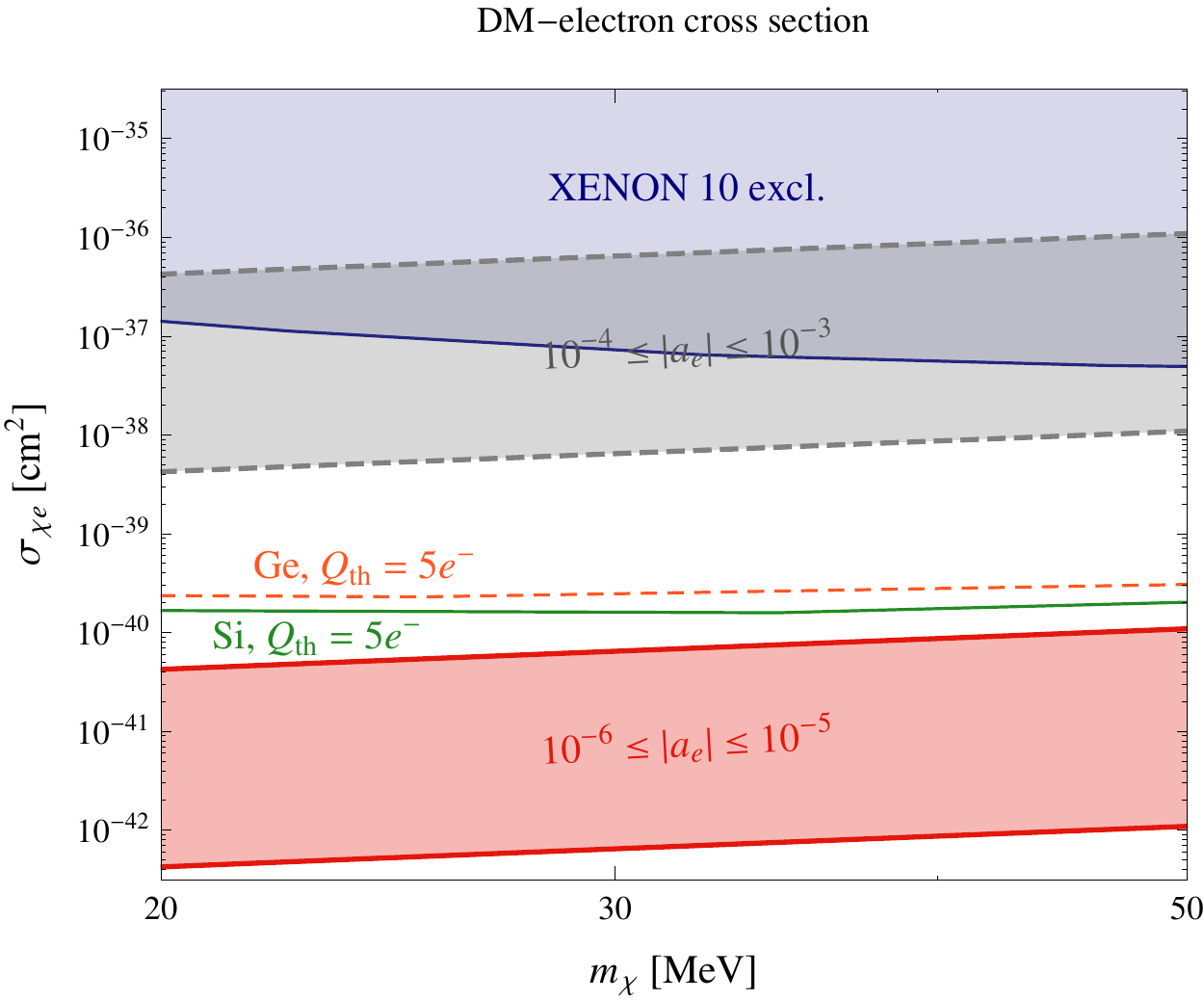}		
		\par\end{centering}
		\caption{\label{fig:DM-electron_sigma}The red shaded regions are the theoretical predictions of $\sigma_{\chi e}$ with $|a_{e}|$ lying in the range of $10^{-6} - 10^{-5}$ for $20~{\rm MeV} < m_{\chi} < 50~{\rm MeV}$. The gray shaded region are the predictions with $|a_{e}|$ lying in the range of $10^{-4} - 10^{-3}$. XENON 10~\cite{Essig:2012yx} excluded the parameter space is represented by purple shaded region. The projected sensitivities of Si and Ge detectors with threshold charges of $5e^{-}$ are represented by the green solid line and the orange dashed line, respectively.}
\end{figure}
It is useful to begin this subsection with discussions in kinematics. We note that $\sigma_{\chi n}^c$ and the predicted $\sigma_{\chi n}$ coincides around $m_{\chi}=50$ MeV. At this mass,  
it is found that $\sigma_{\chi n}^c=10^{-28}$ cm$^2$ while the model predicts $\sigma_{\chi n}=9\times 10^{-29}$ cm$^2$. Hence the overburden of the underground laboratory should be 
less than $1000$ m of standard rock to allow DM of this mass to enter and interact. In the following discussions we assume this is the case and consider $m_{\chi}$ to be between $20$ and $50$ MeV. As said, the lower limit of
$m_{\chi}$ is due to the requirement $m_{\chi}> m_{Z'}$.

We note that the conventional DM direct search looks for the nuclear recoils. However, the nuclear recoil energy, $E_{\rm recoil} = (m_{\chi}v)^2/(2m_A) \approx (m_{\chi}/100~{\rm MeV})^2~(m_A/10~{\rm GeV})^{-1}$~eV, is sub-eV for the MeV-scale DM and is far below the threshold energies in current experiments.\footnote{$v\simeq10^{-3} $ is the DM velocity dispersion in the halo.} Instead of detecting the nuclear recoils, it was suggested that DM-electron scattering can be the DM detection signal~\cite{Bernabei:2007gr,Dedes:2009bk,Kopp:2009et}. The DM-electron cross section $\sigma_{\chi e}$ is given by~\cite{Essig:2011nj}
\begin{equation}\label{sig_xe}
\sigma_{\chi e}=16\pi \alpha_{\rm em}\alpha_d a_e^2\frac{\mu_{\chi e}^2}{M_{Z'}^4}
\end{equation}
where $\mu_{\chi e}$ is the DM-electron reduced mass.
We show the theoretical  predictions to $\sigma_{\chi e}$ and XENON~10 exclusion region in Fig.~\ref{fig:DM-electron_sigma}.  The purple shaded area is excluded by XENON~10 based on the DM-electron scattering and the capability of charge threshold $Q_{\rm th}=10e^-$~\cite{Essig:2012yx}. 
The projected sensitivities for Silicon and Germanium targets with improved $Q_{\rm th}=5e^-$ are presented as green solid line and orange dashed line, respectively~\cite{Essig:2015cda}. The red shaded region is the theoretical predictions of $\sigma_{\chi e}$ corresponding to the range of $10^{-6} < |a_{e}| < 10^{-5}$. 
Our calculation on $\sigma_{\chi e}$ shows that the MeV DM under $Z'$ model
can be tested by future experiments. 

\section{Summary}\label{sec:summary}
Motivated by the possible existence of a new light boson from the experiment of Krasznahorkay \textit{et al}.~\cite{Krasznahorkay:2015iga}, we investigate the $Z'$-portal models. The additional $U(1)$ gauge symmetry may correspond to a hidden charge or a non-hidden charge. The reactor neutrino-electron scattering sets a severe constraint which excludes the parameter space of generic hidden $U(1)$ model. Since the generic hidden $U(1)$ model is disfavored, we are led to consider the non-hidden $U(1)$ portal models. Using the current DM direct search data, we have shown that such models predict $\sigma_{\chi n}$ to be larger than $10^{-27}~{\rm cm^2}$ for $m_{\chi} \geqslant 500$ MeV. Such DM cannot reach the detector located beneath 1000 m of rock and cannot be probed by the DM-nucleon scattering approach. To probe DM lighter than 50 MeV in non-hidden $U(1)$ models, we propose direct searches based upon DM-electron scatterings. The sensitivities of future Si and Ge detectors to $\sigma_{\chi e}$ are given in Fig.~\ref{fig:DM-electron_sigma}. Hence the sensitivities of these detectors to the couplings strength $|a_{e}|$ can be determined accordingly.


\section*{Acknowledgments\label{sec:ackn}}
CSC is supported by Ministry of Science and Technology (MOST), Taiwan, R.O.C. under Grant No.~104-2112-M-001-042-MY3; GLL and YHL are supported by Ministry of Science and Technology, Taiwan under 
Grant No.~104-2112-M-009-021. FRX is supported
by NSFC under Grant No. 11605076
 as well as the Fundamental Research Funds for the Central Universities in China under the Grant No. 21616309. 
 Especially FRX would like to acknowledge the hospitality of Institute of Physics, Academia Sinica, at which part of the work was done.

\end{document}